\documentclass[sigconf]{acmart}

\AtBeginDocument{%
  \providecommand\BibTeX{{%
    \normalfont B\kern-0.5em{\scshape i\kern-0.25em b}\kern-0.8em\TeX}}}

\acmConference{DLP-KDD 2020}{August 24, 2020}{San Diego, California, USA}

\begin{document}

\title{Review Regularized Neural Collaborative Filtering}


\author{Zhimeng Pan}
\affiliation{\institution{University of Utah}}
\email{z.pan@utah.edu}

\author{Wenzheng Tao}
\affiliation{\institution{ University of Utah}}
\email{wztao@cs.utah.edu}

\author{Qingyao Ai}
\affiliation{\institution{ University of Utah}}
\email{aiqy@cs.utah.edu}

\begin{abstract}
	In recent years, text-aware collaborative filtering methods have been proposed to address essential challenges in recommendations such as data sparsity, cold start problem and long-tail distribution. However, many of these text-oriented methods rely heavily on the availability of text information for every user and item, which obviously does not hold in real-world scenarios. Furthermore, specially designed network structures for text processing are highly inefficient for on-line serving and are hard to integrate into current systems.
	 
	In this paper, we propose a flexible neural recommendation framework, named \textbf{R}eview \textbf{R}egularized \textbf{R}ecommendation, short as R3. It consists of a neural collaborative filtering part that focuses on prediction output, and a text processing part that serves as a regularizer. This modular design incorporates text information as richer data sources in the training phase while being highly friendly for on-line serving as it needs no on-the-fly text processing in serving time. Our preliminary results show that by using a simple text processing approach, it could achieve better prediction performance than state-of-the-art text-aware methods.  
\end{abstract}






\maketitle

\section{Introduction}

In the era of Big Data, for most IT related companies, recommendation systems(RecSys) are playing increasingly important roles in boosting their businesses. Currently, Collaborative Filtering(CF) based RecSys and its variants are the most commonly used ones, which stems from a very core assumption that users who purchased(or watched, followed, visited, .etc) similar items shall have similar preference and would enjoy items that share similar features. How to accurately model and therefore measure the similarity between users and items has been a heated research topic over decades. Most modern neural network based algorithms use latent representations(embeddings) to model users and items, and these embeddings are usually computed by factorizing the rating or interaction matrix(Matrix Factorization,MF). MF has been proven to work well for varied recommendation tasks and serves as the building blocks for more complicated RecSys architecture\cite{rendle2019difficulty}. 

However, MF has several major weaknesses. Firstly, it only considers the numerical rating matrix while discarding many useful information like timestamps, order, and texts(reviews, comments or product descriptions). With rich text information in many web services, simply ignoring text data would be a huge waste and there shall be promising potential improvements if we could incorporate these side information into RecSys. Furthermore, since MF is using latent representations, it is hard to explain to the users why such recommendations are made, i.e., the lack of interpretability, which is naturally a consequence of considering solely rating matrix. Finally, due to the extreme sparsity of the rating matrix, MF is easily subject to severe over-fitting, as in \cite{DIN}, authors reported that their model started to over-fit after the first epoch of training. Thus, most MF algorithms apply strong regularizers to the latent factors or only use the simplest interaction function (dot product) to model the relationship between users and items, to avoid over-fitting.  

In recent years, to overcome the aforementioned challenges, researchers in RecSys started trying to utilize those usually discarded side information and come up with RecSys frameworks that are more robust and interpretable \cite{jureHTF, tal2019neural, zhang2014explicit, wang2011collaborative, wang2018explainable}. Currently, most such frameworks choose to incorporate the text information of \textit{user reviews} that is readily available in many web services. They also share the same core idea that user reviews could on the one hand reflect the preference of users and on the other hand characterize the features of products. By forcing, or regularizing the latent factors to represent these preferences and features, RecSys could give explanations on why a recommendation is made, therefore make it more convincing. Meanwhile, by incorporating extra information, one would expect the numerical performance( in terms of recall/precision) of such RecSys would also improve.

There are two major approaches to deal with review text. One is traditional statistical topics modeling like LDA. The other is by using neural approaches like recurrent neural network or neural attention that are widely used in Natural Language Processing community. The statistical approaches are classic, having clear interpretable meanings and cheap to compute, but they are restricted to only a small portion of usage cases ( like topics modeling) due to its intractable posterior, and are not suitable for modeling varied kind of text-related tasks. The main stream of neural approaches are powerful and flexible enough to deal with most kinds of text related tasks. Yet so far, these text-aware neural networks assume the availability of related text for every user and item, which is definitely not the case for real-world applications. Furthermore, these methods use the same architecture for training and inferencing, which means in the inference phase, the models need to complete complicated text processing tasks on-the-fly. This would significantly increase the inference latency, which would be unacceptable for either users and providers.

To address these problems, we propose the R3, a flexible neural recommendation framework, which inherits the two-part structure from \cite{jureHTF}. The main part is a neural collaborative filtering module that takes the user and item embeddings as input and does the actual prediction. The other part is a text-aware regularizer that forces the item embeddings to be meaningful, which is only taking effect in the training phase. With this design, the neural model could utilize as much text information the in training phase, and could also avoid the significant latency increase in the serving phase. Furthermore, unlike highly text-oriented models, it still works where the related text is missing. The whole framework is modular so that we could adopt different scoring functions and regularizers based on the actual need. We show in experiments that a straightforward version of R3 could achieve better prediction performance than state-of-the-art text-aware methods.

The rest of the paper is organized as follows. In section \ref{sec:related} we briefly introduce the foundations of collaborative filtering and related work. In section \ref{sec:alg} we detail the design and training scheme of R3. Experiments are then discussed in section \ref{sec:exp} to demonstrate its effectiveness. In section \ref{sec:ext} we will discuss potential extensions that may further improve the performance or integrate with other recommendation methods. The summary and conclusion will be drawn in the last section.

\section{Back Ground and Related Work}
\label{sec:related}
In a standard setting of latent collaborating filtering model, the predicted rating of user $i$ for an item $j$ is given by 

\begin{equation}
\label{equ:rating}
rec( i,j) = \alpha + \beta_i + \beta_j + rel( \textbf{u}_i, \textbf{v}_j),
\end{equation}
where $\alpha$ is the global offset ( mean) of all ratings, and $\beta_i$ and $\beta_j$ is the local offset of user  $i$ and item $j$ respectively. And $ rel( \textbf{u}_i, \textbf{v}_j)$ is the relevance function that model the interaction ( preference) between user  $i$ and item $j$, taking the latent representation $ \textbf{u}_i$ and $\textbf{v}_j$ as input.

The task of learning would be to find the optimal parameters

\begin{equation}
\label{equ:basic}
\hat{\Theta} =\underset{\Theta}{\mathrm{argmin}} \frac{1}{||D||} \sum_{r_{i,j} \in D} (rec( i, j) - r_{i,j}) ^ 2 + \lambda \Omega( \Theta),
\end{equation}
where $D$ is the training data set, $\Omega(\Theta)$ is the regularization function on parameters $\Theta$, with most commonly used one to be L-2 norm, and $\lambda$ is regularization coefficient.

Classical collaborative filtering methods use dot product as relevance function and L-2 norm as regularization function. The parameters could be either computed through Singular Value Decomposit-ion\cite{rsvd} or sampling\cite{pmf}. Later as the size of data set getting increasingly bigger, mini-batch stochastic gradient descent became the de facto method for finding optimal parameters.

In addition to L-2 norm, in \cite{jureHTF} and \cite{wang2011collaborative}, the authors propose to use LDA as regularizer for item embeddings to overcome data-sparsity and cold-start problem, so as to make the output embedding more explainable.

In \cite{deepConn}, CNN-based text processing was first introduced into collaborative filtering. Users($\textbf{u}_i$) and items($\textbf{v}_j$) are represented by the aggregations of review text embedding they post or received.
Then in \cite{chen2018neural}, neural attention was adopted on top of CNN text processor to learn the usefulness of review and thus improve recommendation quality.

\section{METHODOLOGY}
\label{sec:alg}

\subsection{The R3 Model}
\begin{figure}[h]
	\centering
	\includegraphics[width=\linewidth]{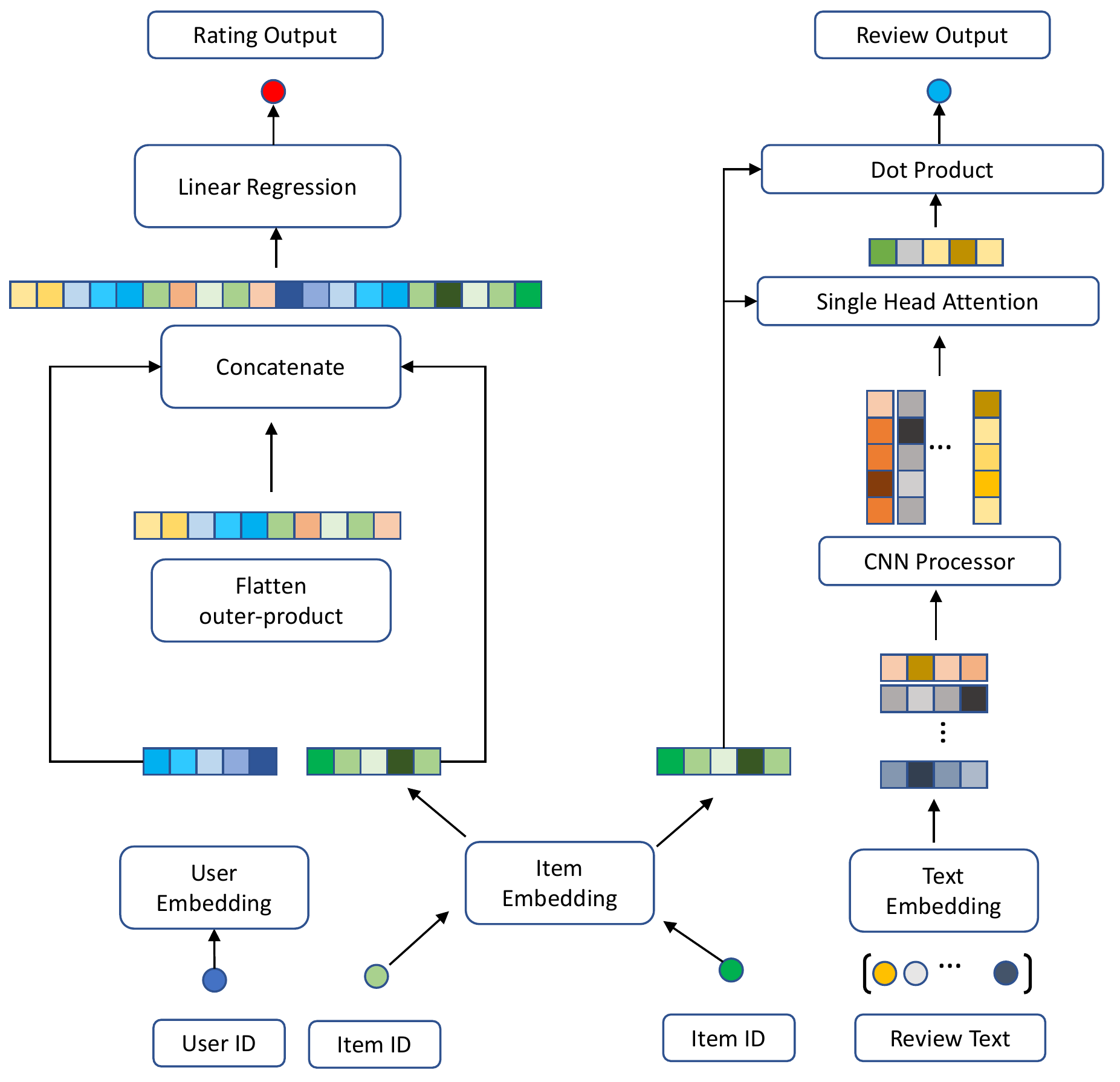}
	\caption{Paradigm of R3. The Left part is the rating part, the right part is the regularization part. Note that both parts share the same item embedding layer}
	\label{fig:network}
\end{figure}

We now detail the design of our proposed framework. The paradigm is shown as figure \ref{fig:network}.

The R3 model consists of two parts, one is the user-item rating part, which serves as the relevance function $rec(i, j)$. The inputs are user ID $i$,  and item ID $j$, which are mapped to dense vectors $\textbf{u}_i$ and $\textbf{v}_j$, of length $K$, by user embedding and item embedding layers respectively. Then we compute and flatten the outer product of $\textbf{u}_i$ and $\textbf{v}_j$, a similar design to \cite{qu2016product}, which was shown to increase the features interactivity. The concatenation of $\textbf{u}_i$, $\textbf{v}_j$ and the flatten outer product is fed into a linear regressor to compute the predicted rating $rec(i, j)$. One could of course choose to use other functions like MLP here, but in this work, we attend to use simple functions. 

The other part is the review regularization part, which is a function $rec_R( j, R_{j,k})$, taking as input the item ID $j$ and one piece of its associated review text $R_{j,k}$. Note that in the training phase, this item ID $j$ does not need to be the same as that in the rating part. One could set another data generator that draws the item ID and review piece separately from ( User ID, Item ID) pairs. The item ID is also mapped to $\textbf{v}_j$ using the same embedding layer as rating part. The padded review text sequence of fixed length $L$ is mapped to a matrix \textbf{H} $\in \mathbb{R}^{L \times D_R}$ using word embeddings generated by word2vector model\cite{word2vec} trained on Google News dataset. During the training we fixed word embeddings and set $D_R = 300$. The review matrix \textbf{H} is then fed into a CNN text processing layer, which consists of $K$ filters of size $w \times D_R$, and outputs a matrix \textbf{M} $\in \mathbb{R}^{L \times K}$. Each filter functions as a simple neural tokenizer to extract phrases up to w-gram. We choose $w=3$ as reported is in \cite{deepConn},and in NLP tasks, phrase length larger then 3 is rarely used. 

To focus on keywords that may characterize product features, we feed $\textbf{v}_j$  and \textbf{M} into a single-head-attention layer as detailed in \cite{attention} with $\textbf{Q}  = \textbf{W}_q\textbf{v}_j$ as query, $\textbf{K} = \textbf{W}_k\textbf{M}$ as keys and $\textbf{V} =\textbf{W}_v\textbf{M}$ as values, where $\textbf{W}_q, \textbf{W}_k,\textbf{W}_v$ are parameter matrices of the layer. The output of this layer would be a vector $\textbf{z}_{j,k}$ of size $K$, which is the weighted sum of values $\textbf{V}$ by attention scores calculated as $softmax( \textbf{Q}\textbf{K}^T)$. The final review output  $rec_R( j, R_{j,k})$ is the dot product of $\textbf{v}_j$ and $\textbf{z}_{j,k}$. The effect of this part is to force the item embedding $\textbf{v}_j$ to be meaningful by extracting and matching key words in review, that may characterize the item features. 

We use both parts' output in the training phase. Yet in the inference phase, only the rating part is needed.

Note that there are not limitations on putting a regularizer on user embeddings. One could definitely do so when it is suitable for their use cases. The reason we only apply the text regularizer on item side is that in our experimental cases, the associated text are reviews for items, which means the text is item-centric, and we naturally expect that the these reviews shall characterize the features of items, and be beneficial for training high quality, meaningful embeddings for items. 

\subsection{Optimization of R3 Model}
For the rating prediction output $ rec(i,j)$, as in conventional regression task, we would like to minimize the mean-square-error(MSE)
\begin{equation}
L_D= \frac{1}{||D||} \sum_{(i,j,r_{i,j}) \in D} ( rec(i,j) - r_{i,j}) ^2,
\end{equation}
where $D$ is the training data set, and $(i, j, r_{i,j})$ is one training sample of user ID, item ID and rating.

As for the regularization output $rec_R( j, R_{j,k})$, one straightforward way is to also treat it as a regression task, minimizing the MSE loss

\begin{equation}
L_R= \frac{1}{||R||} \sum_{(j, R_{j,k}, r_{j,k}) \in R} (rec_R( j, R_{j,k}) - r_{j,k}) ^2,
\end{equation}
where  $(j, R_{j,k}, r_{j,k})$ the a training entry of item ID, one piece of review for this item and the associated rating with this review. $R$ is then the training data set of all these entries, which is normally a subset of D that only contains entries with meaningful reviews. In our setting, meaningful means the review has at least 3 words. 

As pointed out in \cite{kendall2018multi}, in a multi-task learning setting as we here, simply adding these two losses together, maybe with fixed weights, is not an ideal choice. Even in cases where fixed weights might work well, finding the optimal weights is time consuming. Thus in this work, we adopt the proposed solution in \cite{kendall2018multi}. Specifically, the loss function would be uncertainty weighted sum of rating prediction error $L_D$ and review regularization error $L_R$, plus L-2 norm on model parameters

\begin{equation}
Loss = \frac{1}{\sigma_a^2} L_D + \frac{1}{\sigma_b^2} L_R + log \sigma_a\sigma_b+  \lambda||\Theta||^2,
\end{equation} 
where $\sigma_a$ and $\sigma_b$ are self-adaptive parameters representing the degree of uncertainty of the two tasks during training, which would automatically balance the two sub-losses for us. To optimize the final loss, we use mini-batch SGD with Adam optimizer.

\section{Experiments}
\label{sec:exp}

\paragraph{Competing Methods}
To evaluate the effectiveness of our framework, we compare R3 with two kinds of competing methods. One kind is the conventional matrix factorization methods that only account for rating scores. The other kind is those that incorporate text review information into the models, including HFT and DeepCoNN.

\begin{itemize}
	\item \textbf{Stats}. A pure statistical model that calculates the global and local means. No latent factors involved. 
	
	\item \textbf{PMF}. Probabilistic Matrix Factorization\cite{pmf}. Assume latent factors are generated from Gaussian distribution. Equivalent to (\ref{equ:basic})  with regularization function been L-2 norm of latent factors.	
	
	\item \textbf{HFT}. Hidden Factor as Topic proposed in \cite{jureHTF}, which use LDA topical model to regularize the latent factors of items. 
	
	\item \textbf{DeepCoNN}. Deep Cooperative Neural Networks proposed in \cite{deepConn}, jointly models user and item from textual reviews, representing user and item by the embeddings of review words. 
	
\end{itemize}
For HFT\footnote{http://cseweb.ucsd.edu/~jmcauley/code/code\_RecSys13.tar.gz} and DeepCoNN\footnote{https://github.com/chenchongthu/DeepCoNN}, we used the implementation published on open source website. For Stats and PMF we used our own implementation. 

\paragraph{Data Sets and Evaluation Metrics}
In this work we will mainly focus on the numerical rating prediction, evaluated by Root-Mean-Sqrt-Error of the test set, as were done in the original papers of competing methods. We use 3 sub-categories( Movie\&TV, Video Games and Electronics) of the commonly Amazon Reviews Data set 2018 version\footnote{https://nijianmo.github.io/amazon/index.html}. Table \ref{table:1} shows the statistics of the 3 data sets used in experiments. These data sets consist of millions of user review records from Amazon shopping platform from May 1996 to August 2018. Each record contains the reviewer ID, item ID, rating, time-stamp and review text if available. We only used records that are later than Jan-01-2013. As mentioned before, some competing algorithms require that every user and item shall have at least one review. So we preprocessed the data sets by filtering out users and items so that each user and item will have at least 5 associated reviews.(We shall point out that our framework does not have this limitation.)  Then we kept the last rating in time of a user as the held-out test set and the rest as the training set. The splitting ratio is around 1:10. We further split 1/10 of the training set as a validation set for hyper-parameters tuning. 

\begin{table}[h!]
	\centering
	\caption{Data sets statistics after preprocessing( name of category; number of users; number of items; number of ratings/reviews; Sparsity. In this version of datasets, each rating is associated with 5 pieces of reviews. And each user or item has at least 5 ratings/reviews.  )}
	\begin{tabular}{c c c c c} 
		\hline
		Category & \#Users & \#Items & \#Ratings/Reviews & Sparsity\\ [0.5ex] 
		\hline\hline
		Video Games & 29,676 & 22,995 & 286,985 & 0.042\% \\ 
		Movie\&TV & 176,262 & 42,228 & 2,168,289 & 0.029\% \\
		Electronics & 438,659 & 121,073 & 4,566,967 & 0.008\% \\
		\hline
	\end{tabular}
	
	\label{table:1}
\end{table}

\paragraph{Parameter Settings}
There are various hyper-parameters of all these competing methods, and different hyper-parameters settings might lead to significant performance deviation. In order for a fair comparison, we fixed the rank of latent embeddings to be 8, somehow arbitrarily, and tuning other parameters using the validation set. Since HFT and DeepCoNN were also evaluated on Amazon data sets, and delicate hyper-parameters choosing have been done by original authors, we directly used their reported values for structural parameters like the number and width of layers. We did tune the learning rates and early stopping epochs using the validation set, as these parameters are highly sensitive and yet critically important to performance.

\paragraph{Rating Prediction Results}

\begin{table}[h!]
	\caption{Rating prediction results of test set, in Root-Mean-Square-Error}
	\centering
	\begin{tabular}{c c c c c c } 
		\hline
		Category/RMSE & Stats & PMF  &  HFT & DeepCoNN & R3 \\ 
		\hline
		Video Games & 1.321 & 1.168  & 1.095 & 1.102 & \textbf{1.070} \\ 
		
		Movie\&TV & 1.049 & 1.019   & 1.000 & 1.002 & \textbf{0.974} \\
		
		Electronics & 1.213 & 1.256  & 1.184 & 1.193 & \textbf{1.178} \\
		\hline
	\end{tabular}
	
	\label{table:results}
\end{table}

Table \ref{table:results} shows the rating prediction results of all methods involved. For all 3 testing data sets, our proposed framework R3 achieved the lowest RMSE scores. On Movie\&TV, R3's RMSE is significantly lower than the second best models. Comparing to non-text-aware models, text information as an extra training source always improves the model performance. One major difference in utilizing text between R3 and the others ( HFT and DeepCoNN) is that the later methods use the collection of all reviews for one item as the side information in one training entry, no matter what the rating is. While R3 deals with different pieces of reviews separately, so the model could capture the keywords' semantics more accurately depending on the rating. 

We could also note that data sparsity is one of the key factors that affects model performance. When data sparsity is mild, even a simple model like PMF could achieve much better performance than the Stats model. As the sparsity increases, the advantage over Stats model then narrows. For PMF model, it is even worse than Stats on Electronics data sets. This shows the necessity for a strong regularizer like the one used by R3, in scenarios where data is highly sparse.

\paragraph{Inference Speed}

\begin{table}[h!]
	\caption{ Inference speed of selected algorithms, measured in micro seconds per entry, batch size = 1024}
	\centering
	\begin{tabular}{c c c c c c } 
		\hline
		Category/Micro Seconds & DeepCoNN & R3 &\\ 
		\hline
		Video Games   &   199  &0.428 \\ 
		
		Movie\&TV   & 229 & 0.418 \\
		
		Electronics  &  230  & 0.418   \\
		\hline
	\end{tabular}
	
	\label{table:speed}
\end{table}

To demonstrate the efficiency of R3 in inference phase, we compared its inference speed with DeepCoNN \footnote {HFT was written in pure C++ and the inference function is run in single thread. It won't be a fair comparison to include it here}, both implemented using tensorflow. For each category, we measured the inference time by averaging the inference time for 100 batches of size 1024. All experiments were conducted off-line on a server with i7-9700K CPU and 1080Ti GPU. Table \ref{table:speed} shows the speed test result. On all 3 tested categories of data, R3 is around 500 times faster than DeepCoNN. This speed improvement is expected as in inference time, R3 only needs to look up two embeddings ( user and item) then perform an outer product between themselves, followed by a dot product with final weights, with constant time complexity of $\mathcal{O}( K^2)$. On the contrary, DeepCoNN still needs to process the review text ( or at least the their encoded indices), which could vary in length, and look up all the word embeddings, and perform convolution operations on them. These costly operations might be tolerable in training phase but most of the time prohibitive in on-line services. Note that in real-world use cases, there are normally more features besides user and item ids, as such date, time, historical CTR/CVR and so on. As long as these features are easy to fetch and embed, including them into the inference phase model will not increase that much latency. What we are trying to avoid in serving time, is the costly  processing, embedding and  convolution  operations of lengthy text information.

\section{Potential Extensions}
\label{sec:ext}
Benefiting from the modular design, R3 is highly flexible and easy to integrate with other network structures. For the scoring part, besides the user and item embeddings, one could extend the relevance function to include other types of features as inputs like in \cite{wide_deep,dlrm}, or change the relevance function to either linear regression, MLP, or more complicated functions, as long as they work well and are within latency tolerance. For the scoring part, one could choose to use other methods like Neural Topic Modeling in \cite{ntm}, or to regularize users embeddings as \cite{jureHTF} did. These are directions that we could further investigate in the future.  
\section{Conclusion}
In this paper, we propose a flexible neural recommendation framework R3, which utilizes review text as regularizer. Our experiments show that compared to state-of-the-art text-aware methods, it could achieve better prediction performance while preserving highly efficient serving speed. There are many potential directions to further integrate into this framework to further improve its effectiveness.   

\bibliographystyle{ACM-Reference-Format}
\bibliography{NRRCF}


\begin{thebibliography}{18}


\ifx \showCODEN    \undefined \def \showCODEN     #1{\unskip}     \fi
\ifx \showDOI      \undefined \def \showDOI       #1{#1}\fi
\ifx \showISBNx    \undefined \def \showISBNx     #1{\unskip}     \fi
\ifx \showISBNxiii \undefined \def \showISBNxiii  #1{\unskip}     \fi
\ifx \showISSN     \undefined \def \showISSN      #1{\unskip}     \fi
\ifx \showLCCN     \undefined \def \showLCCN      #1{\unskip}     \fi
\ifx \shownote     \undefined \def \shownote      #1{#1}          \fi
\ifx \showarticletitle \undefined \def \showarticletitle #1{#1}   \fi
\ifx \showURL      \undefined \def \showURL       {\relax}        \fi
\providecommand\bibfield[2]{#2}
\providecommand\bibinfo[2]{#2}
\providecommand\natexlab[1]{#1}
\providecommand\showeprint[2][]{arXiv:#2}

\bibitem[\protect\citeauthoryear{Cao, Li, Liu, Li, and Ji}{Cao
  et~al\mbox{.}}{2015}]%
        {ntm}
\bibfield{author}{\bibinfo{person}{Ziqiang Cao}, \bibinfo{person}{Sujian Li},
  \bibinfo{person}{Yang Liu}, \bibinfo{person}{Wenjie Li}, {and}
  \bibinfo{person}{Heng Ji}.} \bibinfo{year}{2015}\natexlab{}.
\newblock \showarticletitle{A novel neural topic model and its supervised
  extension}. In \bibinfo{booktitle}{\emph{Twenty-Ninth AAAI Conference on
  Artificial Intelligence}}.
\newblock


\bibitem[\protect\citeauthoryear{Chen, Zhang, Liu, and Ma}{Chen
  et~al\mbox{.}}{2018}]%
        {chen2018neural}
\bibfield{author}{\bibinfo{person}{Chong Chen}, \bibinfo{person}{Min Zhang},
  \bibinfo{person}{Yiqun Liu}, {and} \bibinfo{person}{Shaoping Ma}.}
  \bibinfo{year}{2018}\natexlab{}.
\newblock \showarticletitle{Neural attentional rating regression with
  review-level explanations}. In \bibinfo{booktitle}{\emph{Proceedings of the
  2018 World Wide Web Conference}}. \bibinfo{pages}{1583--1592}.
\newblock


\bibitem[\protect\citeauthoryear{Cheng, Koc, Harmsen, Shaked, Chandra, Aradhye,
  Anderson, Corrado, Chai, Ispir, et~al\mbox{.}}{Cheng et~al\mbox{.}}{2016}]%
        {wide_deep}
\bibfield{author}{\bibinfo{person}{Heng-Tze Cheng}, \bibinfo{person}{Levent
  Koc}, \bibinfo{person}{Jeremiah Harmsen}, \bibinfo{person}{Tal Shaked},
  \bibinfo{person}{Tushar Chandra}, \bibinfo{person}{Hrishi Aradhye},
  \bibinfo{person}{Glen Anderson}, \bibinfo{person}{Greg Corrado},
  \bibinfo{person}{Wei Chai}, \bibinfo{person}{Mustafa Ispir}, {et~al\mbox{.}}}
  \bibinfo{year}{2016}\natexlab{}.
\newblock \showarticletitle{Wide \& deep learning for recommender systems}. In
  \bibinfo{booktitle}{\emph{Proceedings of the 1st workshop on deep learning
  for recommender systems}}. \bibinfo{pages}{7--10}.
\newblock


\bibitem[\protect\citeauthoryear{Kendall, Gal, and Cipolla}{Kendall
  et~al\mbox{.}}{2018}]%
        {kendall2018multi}
\bibfield{author}{\bibinfo{person}{Alex Kendall}, \bibinfo{person}{Yarin Gal},
  {and} \bibinfo{person}{Roberto Cipolla}.} \bibinfo{year}{2018}\natexlab{}.
\newblock \showarticletitle{Multi-task learning using uncertainty to weigh
  losses for scene geometry and semantics}. In
  \bibinfo{booktitle}{\emph{Proceedings of the IEEE conference on computer
  vision and pattern recognition}}. \bibinfo{pages}{7482--7491}.
\newblock


\bibitem[\protect\citeauthoryear{Le and Mikolov}{Le and Mikolov}{2014}]%
        {word2vec}
\bibfield{author}{\bibinfo{person}{Quoc Le} {and} \bibinfo{person}{Tomas
  Mikolov}.} \bibinfo{year}{2014}\natexlab{}.
\newblock \showarticletitle{Distributed representations of sentences and
  documents}. In \bibinfo{booktitle}{\emph{International conference on machine
  learning}}. \bibinfo{pages}{1188--1196}.
\newblock


\bibitem[\protect\citeauthoryear{McAuley and Leskovec}{McAuley and
  Leskovec}{2013}]%
        {jureHTF}
\bibfield{author}{\bibinfo{person}{Julian McAuley} {and} \bibinfo{person}{Jure
  Leskovec}.} \bibinfo{year}{2013}\natexlab{}.
\newblock \showarticletitle{Hidden factors and hidden topics: understanding
  rating dimensions with review text}. In \bibinfo{booktitle}{\emph{Proceedings
  of the 7th ACM conference on Recommender systems}}.
  \bibinfo{pages}{165--172}.
\newblock


\bibitem[\protect\citeauthoryear{Mnih and Salakhutdinov}{Mnih and
  Salakhutdinov}{2008}]%
        {pmf}
\bibfield{author}{\bibinfo{person}{Andriy Mnih} {and} \bibinfo{person}{Russ~R
  Salakhutdinov}.} \bibinfo{year}{2008}\natexlab{}.
\newblock \showarticletitle{Probabilistic matrix factorization}. In
  \bibinfo{booktitle}{\emph{Advances in neural information processing
  systems}}. \bibinfo{pages}{1257--1264}.
\newblock


\bibitem[\protect\citeauthoryear{Naumov, Mudigere, Shi, Huang, Sundaraman,
  Park, Wang, Gupta, Wu, Azzolini, et~al\mbox{.}}{Naumov et~al\mbox{.}}{2019}]%
        {dlrm}
\bibfield{author}{\bibinfo{person}{Maxim Naumov}, \bibinfo{person}{Dheevatsa
  Mudigere}, \bibinfo{person}{Hao-Jun~Michael Shi}, \bibinfo{person}{Jianyu
  Huang}, \bibinfo{person}{Narayanan Sundaraman}, \bibinfo{person}{Jongsoo
  Park}, \bibinfo{person}{Xiaodong Wang}, \bibinfo{person}{Udit Gupta},
  \bibinfo{person}{Carole-Jean Wu}, \bibinfo{person}{Alisson~G Azzolini},
  {et~al\mbox{.}}} \bibinfo{year}{2019}\natexlab{}.
\newblock \showarticletitle{Deep learning recommendation model for
  personalization and recommendation systems}.
\newblock \bibinfo{journal}{\emph{arXiv preprint arXiv:1906.00091}}
  (\bibinfo{year}{2019}).
\newblock


\bibitem[\protect\citeauthoryear{Paterek}{Paterek}{2007}]%
        {rsvd}
\bibfield{author}{\bibinfo{person}{Arkadiusz Paterek}.}
  \bibinfo{year}{2007}\natexlab{}.
\newblock \showarticletitle{Improving regularized singular value decomposition
  for collaborative filtering}. In \bibinfo{booktitle}{\emph{Proceedings of KDD
  cup and workshop}}, Vol.~\bibinfo{volume}{2007}. \bibinfo{pages}{5--8}.
\newblock


\bibitem[\protect\citeauthoryear{Qu, Cai, Ren, Zhang, Yu, Wen, and Wang}{Qu
  et~al\mbox{.}}{2016}]%
        {qu2016product}
\bibfield{author}{\bibinfo{person}{Yanru Qu}, \bibinfo{person}{Han Cai},
  \bibinfo{person}{Kan Ren}, \bibinfo{person}{Weinan Zhang},
  \bibinfo{person}{Yong Yu}, \bibinfo{person}{Ying Wen}, {and}
  \bibinfo{person}{Jun Wang}.} \bibinfo{year}{2016}\natexlab{}.
\newblock \showarticletitle{Product-based neural networks for user response
  prediction}. In \bibinfo{booktitle}{\emph{2016 IEEE 16th International
  Conference on Data Mining (ICDM)}}. IEEE, \bibinfo{pages}{1149--1154}.
\newblock


\bibitem[\protect\citeauthoryear{Rendle, Zhang, and Koren}{Rendle
  et~al\mbox{.}}{2019}]%
        {rendle2019difficulty}
\bibfield{author}{\bibinfo{person}{Steffen Rendle}, \bibinfo{person}{Li Zhang},
  {and} \bibinfo{person}{Yehuda Koren}.} \bibinfo{year}{2019}\natexlab{}.
\newblock \showarticletitle{On the difficulty of evaluating baselines: A study
  on recommender systems}.
\newblock \bibinfo{journal}{\emph{arXiv preprint arXiv:1905.01395}}
  (\bibinfo{year}{2019}).
\newblock


\bibitem[\protect\citeauthoryear{Tal, Liu, Huang, Yu, and Aljbawi}{Tal
  et~al\mbox{.}}{2019}]%
        {tal2019neural}
\bibfield{author}{\bibinfo{person}{Omer Tal}, \bibinfo{person}{Yang Liu},
  \bibinfo{person}{Jimmy Huang}, \bibinfo{person}{Xiaohui Yu}, {and}
  \bibinfo{person}{Bushra Aljbawi}.} \bibinfo{year}{2019}\natexlab{}.
\newblock \showarticletitle{Neural Attention Frameworks for Explainable
  Recommendation}.
\newblock \bibinfo{journal}{\emph{IEEE Transactions on Knowledge and Data
  Engineering}} (\bibinfo{year}{2019}).
\newblock


\bibitem[\protect\citeauthoryear{Vaswani, Shazeer, Parmar, Uszkoreit, Jones,
  Gomez, Kaiser, and Polosukhin}{Vaswani et~al\mbox{.}}{2017}]%
        {attention}
\bibfield{author}{\bibinfo{person}{Ashish Vaswani}, \bibinfo{person}{Noam
  Shazeer}, \bibinfo{person}{Niki Parmar}, \bibinfo{person}{Jakob Uszkoreit},
  \bibinfo{person}{Llion Jones}, \bibinfo{person}{Aidan~N Gomez},
  \bibinfo{person}{{\L}ukasz Kaiser}, {and} \bibinfo{person}{Illia
  Polosukhin}.} \bibinfo{year}{2017}\natexlab{}.
\newblock \showarticletitle{Attention is all you need}. In
  \bibinfo{booktitle}{\emph{Advances in neural information processing
  systems}}. \bibinfo{pages}{5998--6008}.
\newblock


\bibitem[\protect\citeauthoryear{Wang and Blei}{Wang and Blei}{2011}]%
        {wang2011collaborative}
\bibfield{author}{\bibinfo{person}{Chong Wang} {and} \bibinfo{person}{David~M
  Blei}.} \bibinfo{year}{2011}\natexlab{}.
\newblock \showarticletitle{Collaborative topic modeling for recommending
  scientific articles}. In \bibinfo{booktitle}{\emph{Proceedings of the 17th
  ACM SIGKDD international conference on Knowledge discovery and data mining}}.
  \bibinfo{pages}{448--456}.
\newblock


\bibitem[\protect\citeauthoryear{Wang, Wang, Jia, and Yin}{Wang
  et~al\mbox{.}}{2018}]%
        {wang2018explainable}
\bibfield{author}{\bibinfo{person}{Nan Wang}, \bibinfo{person}{Hongning Wang},
  \bibinfo{person}{Yiling Jia}, {and} \bibinfo{person}{Yue Yin}.}
  \bibinfo{year}{2018}\natexlab{}.
\newblock \showarticletitle{Explainable recommendation via multi-task learning
  in opinionated text data}. In \bibinfo{booktitle}{\emph{The 41st
  International ACM SIGIR Conference on Research \& Development in Information
  Retrieval}}. \bibinfo{pages}{165--174}.
\newblock


\bibitem[\protect\citeauthoryear{Zhang, Lai, Zhang, Zhang, Liu, and Ma}{Zhang
  et~al\mbox{.}}{2014}]%
        {zhang2014explicit}
\bibfield{author}{\bibinfo{person}{Yongfeng Zhang}, \bibinfo{person}{Guokun
  Lai}, \bibinfo{person}{Min Zhang}, \bibinfo{person}{Yi Zhang},
  \bibinfo{person}{Yiqun Liu}, {and} \bibinfo{person}{Shaoping Ma}.}
  \bibinfo{year}{2014}\natexlab{}.
\newblock \showarticletitle{Explicit factor models for explainable
  recommendation based on phrase-level sentiment analysis}. In
  \bibinfo{booktitle}{\emph{Proceedings of the 37th international ACM SIGIR
  conference on Research \& development in information retrieval}}.
  \bibinfo{pages}{83--92}.
\newblock


\bibitem[\protect\citeauthoryear{Zheng, Noroozi, and Yu}{Zheng
  et~al\mbox{.}}{2017}]%
        {deepConn}
\bibfield{author}{\bibinfo{person}{Lei Zheng}, \bibinfo{person}{Vahid Noroozi},
  {and} \bibinfo{person}{Philip~S Yu}.} \bibinfo{year}{2017}\natexlab{}.
\newblock \showarticletitle{Joint deep modeling of users and items using
  reviews for recommendation}. In \bibinfo{booktitle}{\emph{Proceedings of the
  Tenth ACM International Conference on Web Search and Data Mining}}.
  \bibinfo{pages}{425--434}.
\newblock


\bibitem[\protect\citeauthoryear{Zhou, Zhu, Song, Fan, Zhu, Ma, Yan, Jin, Li,
  and Gai}{Zhou et~al\mbox{.}}{2018}]%
        {DIN}
\bibfield{author}{\bibinfo{person}{Guorui Zhou}, \bibinfo{person}{Xiaoqiang
  Zhu}, \bibinfo{person}{Chenru Song}, \bibinfo{person}{Ying Fan},
  \bibinfo{person}{Han Zhu}, \bibinfo{person}{Xiao Ma},
  \bibinfo{person}{Yanghui Yan}, \bibinfo{person}{Junqi Jin},
  \bibinfo{person}{Han Li}, {and} \bibinfo{person}{Kun Gai}.}
  \bibinfo{year}{2018}\natexlab{}.
\newblock \showarticletitle{Deep interest network for click-through rate
  prediction}. In \bibinfo{booktitle}{\emph{Proceedings of the 24th ACM SIGKDD
  International Conference on Knowledge Discovery \& Data Mining}}.
  \bibinfo{pages}{1059--1068}.
\newblock


\end{thebibliography}

\end{document}